\documentclass[journal]{IEEEtran}
\usepackage{xr,graphicx,times,amsmath,amssymb,algorithmic,algorithm,multirow,color}
\usepackage[normalem]{ulem}
\usepackage{booktabs}
\newtheorem{theorem}{\bf Theorem}

\usepackage{color}

\begin{document}

 \title{The N-Player Trust Game and its Replicator Dynamics}

\author{Hussein Abbass, Garrison Greenwood, and Eleni Petraki
 \thanks{Hussein Abbass is with the School of Engineering \& IT, University of New South Wales,
 Canberra-Australia, (e-mail: h.abbass@adfa.edu.au).}
 \thanks{Garrison Greenwood is with the Electrical and Computer Engineering Department, Portland State University, Oregon - USA
 (e-mail: greenwd@pdx.edu).}
 \thanks{Eleni Petraki is with the Faculty of Arts and Design, University of Canberra Canberra, Australia
 (e-mail: Eleni.Petraki@canberra.edu.au).}
}

\maketitle

\begin{abstract}
Trust is a fundamental concept that underpins the coherence and
resilience of social systems and shapes human behavior. Despite
the importance of trust as a social and psychological concept, the
concept has not gained much attention from evolutionary game
theorists. In this paper, an N-player trust-based social dilemma
game is introduced. While the theory shows that a society with no
untrustworthy individuals would yield maximum wealth to both the
society as a whole and the individuals in the long run,
evolutionary dynamics show this ideal situation is reached only in
a special case when the initial population contains no
untrustworthy individuals. When the initial population consists of
even the slightest number of untrustworthy individuals, the society
converges to zero trusters, with many untrustworthy individuals.
The promotion of trust is an uneasy task, despite the fact that a
combination of trusters and trustworthy trustees is the most
rational and optimal social state. This paper presents the game
and results of replicator dynamics in a hope that researchers in
evolutionary games see opportunities in filling this critical gap
in the literature.
\end{abstract}


\IEEEpeerreviewmaketitle

\section{Introduction}

\IEEEPARstart{T}{rust} is the glue of a social
system~\cite{luhmann1979trust32}. Despite its vital role in the
society, the concept is absent from the evolutionary computation
literature. When compared to the large number of papers published
on the iterated prisoner
dilemma~\cite{fogel1993evolving,darwen1997speciation,ishibuchi2005evolution,Ashl1,mittal2009optimal,li2014effect},
there has not been any publication on trust. The contributions of
this paper are three-fold. First, it aims to encourage more work
on trust in the evolutionary computation and evolutionary game
theory research areas. Second, it introduces a novel N-player
trust game that assists researchers in understanding and analyzing
the concept of trust using evolutionary game theory. The game is a
social dilemma and generalizes the concept of trust, which is
normally modelled as a sequential game, to a population of players
that can play the game concurrently. Third, it presents the first
theoretical analysis of the dynamics of the game using replicator
dynamics.

A recent review on trust in social and psychological
literature~\cite{petraki2014trust0} demonstrated that there are
abundant studies on the roles of
trust~\cite{deutsch1962cooperation12,deutsch1977resolution13,luhmann1979trust32}
and its implications for social and human
systems~\cite{fehr2003nature,gambetta2000can14,grodzinsky2014developing19,reina2000trust23},
including ethical considerations~\cite{baier1986trust7}
surrounding studies of trust and the means for influencing and
shaping trust~\cite{brooks2014m10,mayer1995integrative33}.
Managers perceive how an understanding of how trust is formed can
create opportunities to develop loyal
customers~\cite{ball2004role8} and improve relationships among
employees and
management~\cite{helliwell2011well20,wells2001trust43,spector2004trust41}.
Individuals perceive trust as a
vulnerability~\cite{mayer1995integrative33} in which one person
exposes himself to another by relying on the other person to
make a decision on his/her behalf, as an
opportunity~\cite{morrison1997employees36} to create favorable
outcomes, and as a source of unwanted
uncertainty~\cite{giffin1967contribution24} that creates a
relationship in which the trustee has power over the trusting
party.

In the social and psychological literature, trust is described as
playing two important roles. First, sociologically, it acts as a
complexity-reduction mechanism~\cite{luhmann1979trust32}, allowing
individuals with limited cognitive capacities to manage the
complex world they live in. Luhmann~\cite{luhmann1979trust32}
showed that trust creates a positive feedback loop in a social
system. As individuals use trust to manage complexity,
relationships emerge, a process that increases complexity and thus
creates more reinforcement and opportunities for trust to spread.

Second, psychologically, it acts as an ambiguity- and
uncertainty-reduction mechanism for
individuals~\cite{delgado2005perceptions11,deutsch1977resolution13}.
Deutsch~\cite{deutsch1977resolution13} argues that a trusting
situation occurs when a truster perceives that a situation has one
of two potential outcomes, where one is perceived to have a
negative valency of greater absolute value than the positive
valency attributed to the second. However, which outcome will
occur depends on the trustee. If the truster chooses to proceed,
the truster is said to trust the trustee; otherwise, the truster
distrusts the trustee. Context influences the perceived levels of
both trustworthiness and risk; as the risk of trusting changes
from one context to another, the decision to trust changes as
well. Therefore, trust and risk are tightly coupled concepts.

Trust games are sequential in nature. The truster must first
decide whether to trust the trustee. If the decision is to trust
the trustee, the trustee must then decide whether to be
trustworthy or not. Each decision can be a binary decision or be
considered as part of a continuum representing the degree of
either trust or trustworthiness. Although the proposed game relies
on binary decision-making, it can easily be generalized to include
non-binary decisions. Neuroeconomic experiments use two main forms
of trust
games~\cite{adolphs2002trust5,delgado2005perceptions11,giffin1967contribution24}.
We will call them TG1 and TG2 for trust game 1 and 2,
respectively.

TG1~\cite{krueger2007neural26} is a non-zero-sum game, and the
truster must choose whether to trust the trustee. If the truster
chooses not to trust the trustee, both players receive \$5. If the
truster chooses to trust the trustee, then the trustee must decide
whether to be trustworthy or not. If the trustee chooses to be
trustworthy, the truster and trustee receive \$10 and \$15,
respectively. If the trustee chooses to defect, the truster gets
nothing, whereas the trustee gets \$25.

\begin{figure}[!t]
\centering
\includegraphics[width=3in,height=2.2in]{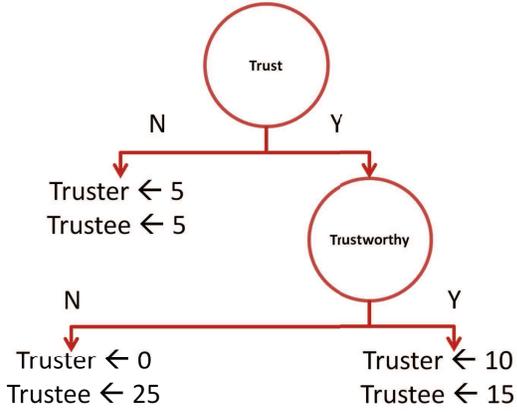}
\caption{Game tree for a 2-player TG1.} \label{TrustingF1}
\end{figure}

In this game, a rational truster would be indifferent to the two
choices under the assumption that the trustee's probability of
being trustworthy is 0.5. Therefore, the expected return for the
truster, regardless of the decision to trust or not, is \$5. The
trustee, however, benefits more if the truster chooses to trust
the trustee. In this case, in a one-off situation, a rational
trustee would choose to defect to maximize his or her own return.
However, over repeated iterations, the trustee has an incentive to
be trustworthy because being so will yield maximum returns to both
the trustee and truster. The game tree for this game can be seen
in Figure~\ref{TrustingF1}.

\begin{figure}[!t]
\centering
\includegraphics[width=3in,height=2.2in]{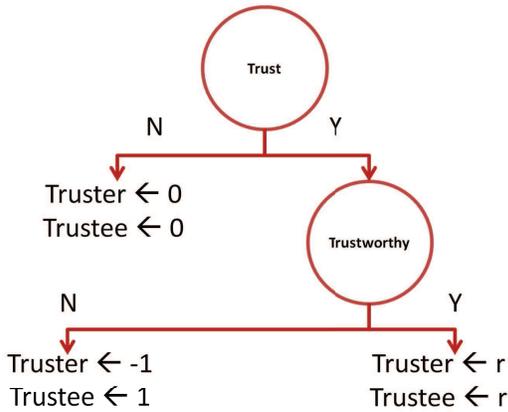}
\caption{Game tree for a general 2-player TG1G.}
\label{TrustingF1G}
\end{figure}

TG1 can be generalized, we call it TG1G, to the game tree shown in
Figure~\ref{TrustingF1G}. An early version of this general form
was presented in~\cite{dasgupta2000trust} and the general version
was then analyzed in~\cite{masuda2012coevolution}.

TG2~\cite{adolphs2002trust5} is also a non-zero-sum game. Both the
truster and trustee start with an endowment of \$12. The truster
makes the first move by deciding how much money to transfer to the
trustee. The money is tripled on the way to the trustee. The
trustee then must decide how much money to transfer back to the
truster. If X is the amount of money the truster sent to the
trustee and Y is the amount of money the trustee sends back to the
truster, then the truster will end up with a balance of 12 - X +
Y¸ while the trustee ends up with a balance of 12 + 3X - Y. Both
the truster and trustee end up with the same balance when Y = 2X.
In this game, X is a measure of trust and Y is a measure of
trustworthiness. This game is represented in the following table.

\begin{table}[!t]
 \renewcommand{\arraystretch}{1.3}
 \caption{Utility matrix for a 2-player trust game.}\label{TrustingG2}
 \newcommand{\tabincell}[2]{\begin{tabular}{@{}#1@{}}#2\end{tabular}}
 \centering
 \begin{tabular}{|l|c|c|}
 \hline
 & {\bfseries Truster} & {\bfseries Trustee} \\
 \hline
 {\bfseries Starting Balance} & 12 & 12 \\
 {\bfseries Pay} & $X$ & $Y$ \\
 {\bfseries Receive} & $Y$ & $3X$ \\
 {\bfseries Net Wealth} & $12-X+Y$ & $12+3X-Y$ \\
 \hline
\end{tabular}
\end{table}

Neither TG1 nor TG2 create a complete social dilemma. If the
truster chooses to trust the trustee, then the total wealth,
independent of whether the trustee is trustworthy or not, is \$25
in TG1 and $24 + 2X$ in TG2. The social dilemma, therefore, is
reliant only on the truster's decision. However, in TG1G, the game
is a social dilemma when $0<r<1$.

\section{N-PLAYER TRUST GAME}

Assume $N$ players. Each player must make two decisions in
advance. The first is to decide whether to be trustworthy or not.
The second is to decide whether to govern (be a governor or a
trustee) or be governed (be a citizen or a truster). We will
denote the former as $G$ and the latter as $C$. Assume the number
of players that decided to be type $C$ is $x_1$ and the number of
players that decided to be type $G$ is $x_2+x_3$; such that, $N =
x_1 + x_2 + x_3$.

In real-world settings, the role of the truster and trustee are
defined \textit{a priori}. For example, an investor chooses to invest,
while a financial planning advisor chooses to become one before
the two actors decide to enter into a trusting relationship. This
type of decision we refer to as a social choice. The second
type of decision normally decided \textit{a priori} as well is
whether the trustee will be trustworthy or not. The reason it is
`normally' decided \textit{a priori} is that it largely depends on the
behavioral attitude and value system of the trustee. We
acknowledge that context influences this behavioral attribute but
we conjecture that this decision is fundamentally a core
behavioral attribute of the agent. We call this decision a
behavioral choice.

In summary, it is reasonable to assume that the two types of
decisions related to social and behavioral choices can be done
\textit{a priori}.

A player of type $C$ pays $tv$ to the government, where $tv$
denotes the trusted value. The dynamics of the game is in general
independent of the value of $tv$, which can be set to 1. However,
we maintain $tv$ to allow flexibility in adopting the game to
different contexts. With $x_1$ players of type $C$, the total
money that is sent to the government is $x_1 \cdot tv$. Each
player of type $G$ receives $(x_1 \cdot tv)/(x_2+x_3)$. Assume
$x_2$ players of type $G$ decide to be trustworthy, while $x_3$
decide to not be trustworthy. A player in the $x_2$ population
returns to the citizens a multiplier of $R1$ of what was received
and keeps the same amount for himself, with $R1 > 1$. A player in
the $x_3$ population returns nothing to the citizens and keeps for
himself a multiplier of $R2$ of what was received, where $R1 < R2
< 2R1$. The payoff matrix for this game can then be represented as
shown in Table~\ref{TrustingT1} with the following constraints:

 \begin{equation} 1<R1<R2<2R1 \end{equation}
 \begin{equation} N=x_1 + x_2 + x_3 \end{equation}

\begin{table*}[!t]
 \renewcommand{\arraystretch}{1.3}
 \caption{Utility matrix for a N-player trust game.}\label{TrustingT1}
 \newcommand{\tabincell}[2]{\begin{tabular}{@{}#1@{}}#2\end{tabular}}
 \centering
 \begin{tabular}{lccc}
 \toprule
 & {\bfseries Player in the $x_1$ population} &  {\bfseries Player in the $x_2$ population} & {\bfseries Player in the $x_3$ population} \\ \midrule
 Pay & $tv$ & $R1\cdot tv\cdot\frac{x_1}{x_2+x_3}$ & 0\\
 Receive & $R1\cdot tv\cdot\frac{x_2}{x_2+x_3}$ & $2\cdot R1\cdot tv\cdot \frac{x_1}{x_2+x_3}$ & $R2\cdot tv\cdot \frac{x_1}{x_2+x_3}$ \\
 Net Wealth & $tv\cdot (R1\cdot\frac{x_2}{x_2+x_3}-1)$ & $R1\cdot tv\cdot\frac{x_1}{x_2+x_3}$ & $R2\cdot tv\cdot\frac{x_1}{x_2+x_3}$ \\
 \bottomrule
\end{tabular}
\end{table*}

\begin{figure}[!t]
\centering
\includegraphics[width=3in,height=2.2in]{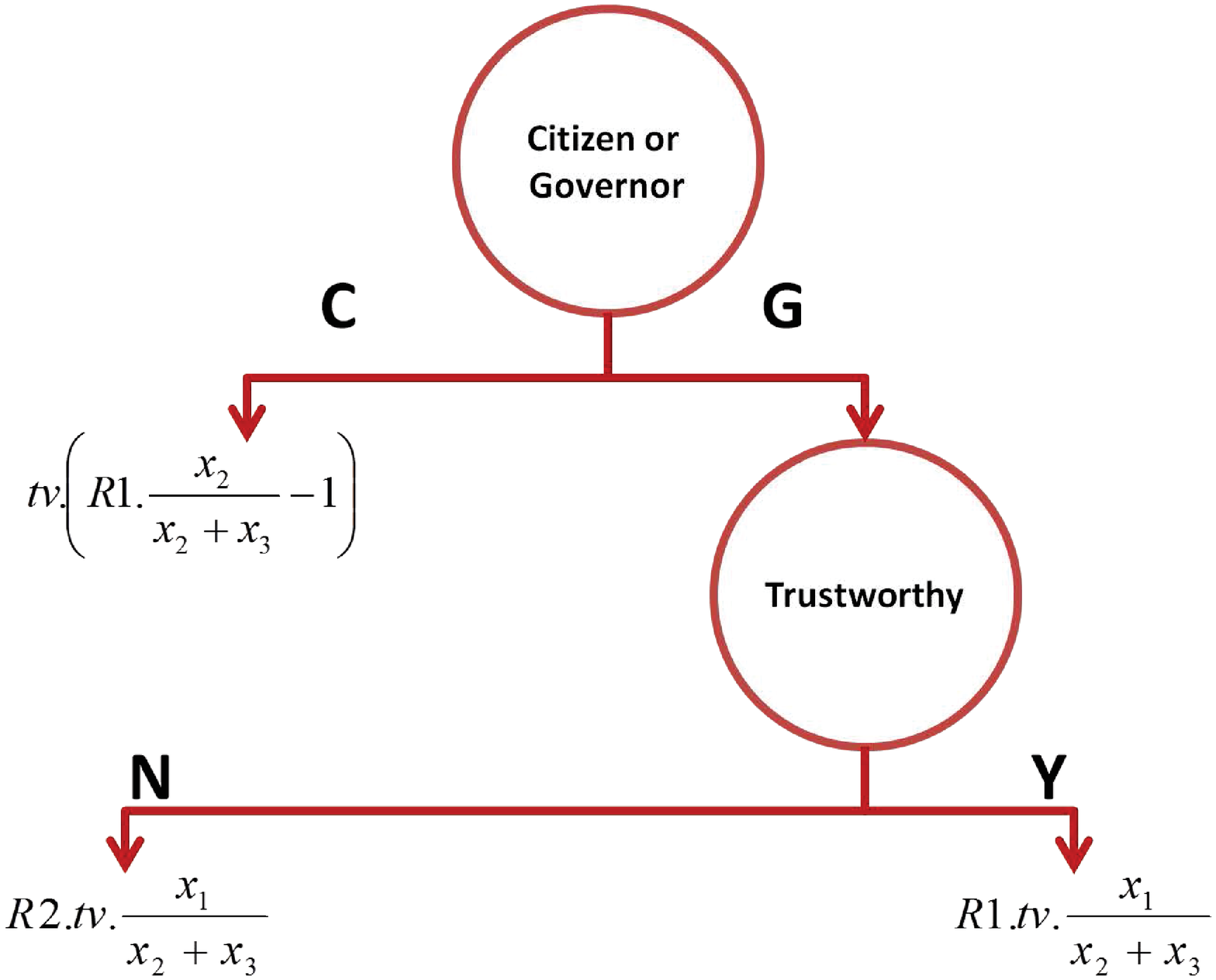}
\caption{Game tree for the proposed N-player trust game.}
\label{TrustingF2}
\end{figure}

The game tree is presented in Figure~\ref{TrustingF2} for an
arbitrary number of players.

\begin{theorem}
$x_3=N$ is a Nash Equilibrium.
\end{theorem}

\vspace{2ex}

\textit{Proof.} The order of the net individual wealth in the
table from left to right is a monotonically increasing sequence. A
player deciding to play according to any strategy on the
right-hand side will yield a better net individual wealth than any
of the strategies on the left-hand side. The best net individual
wealth occurs for players in $x_3$. Any rational player will
choose to be a governor and untrustworthy. Thus, the Nash
equilibrium occurs when all players choose to be governors and
untrustworthy. $\blacksquare$

The total combined wealth $CW$ of the population is

\begin{equation}
CW=\left\{
\begin{tabular}{ll}
 $x_1\cdot tv\left( \frac{2\cdot R1 \cdot x_2}{x_2+x_3} + \frac{R2 \cdot x_3}{x_2+x_3} -1 \right)$ & if $x_2+x_3 \ge 1$ \\
 $0$ & \textrm{otherwise} \\
\end{tabular}
\right.
\label{cw-eqn}
\end{equation}

\vspace{1ex}

\begin{theorem}
The maximum of $CW$ occurs when $x_2=1$ and $x_3=0$.
\end{theorem}

\vspace{2ex}

\textit{Proof.}
Given the constraints on $R1$ and $R2$, $CW$ has a lower bound of
zero regardless of the value of $x_2+x_3$. Therefore, to maximize
$CW$, we only need to focus on the case when $x_2+x_3 \ge 1$.
Substituting $x_3=\left(x_2+x_3\right)-x_2$ into Eq.~(\ref{cw-eqn}),

\begin{eqnarray*}
CW & = & x_1 \cdot tv\cdot\left( \frac{2R1\cdot x_2}{x_2+x_3}-1+R2-\frac{R2x_2}{x_2+x_3} \right) \\
& = & x_1\cdot tv\cdot\left( \frac{\left( 2R1-R2 \right) \cdot x_2}{x_2+x_3}  +(R2-1) \right) \\
& = & (N-(x_2+x_3))\cdot tv\cdot \\
& & \hspace{0.49in} \left( \frac{\left( 2R1-R2 \right) \cdot x_2}{x_2+x_3} +(R2-1) \right)
 \end{eqnarray*}

\vspace{1ex}

Recall that $R2>(2 R1-R2) > 0$, $N$ is a constant, and $x_2+x_3
\ge 1$, the maximum of the function will occur when $x_2+x_3$ is
at a minimum ($x_2+x_3$ appears twice, once with a negative sign
in the numerator and once in the denominator) while $x_3=0$ is the
maximum for $\frac{x_2}{x_2+x_3}$; therefore, $x_2+x_3=1 \ \
\Rightarrow \ \ x_2=1$. $\blacksquare$

\vspace{4ex}

\begin{theorem}\label{Th3} $x_1 = N-1$ and $x_2=1$ is Pareto Optimal. \end{theorem}

\vspace{2ex}

\textit{Proof.} The single $x_2$ player can switch to an $x_3$
player or an $x_1$ player; either change reduces the $x_1$ player
payoff to zero. All investments are split among the $x_2$ and
$x_3$ players so any $x_1$ player that changes to an $x_2$ or
$x_3$ player reduces the payoff to the single $x_2$ player. The
proof follows that any player unilaterally switching roles reduces
the payoff to an $x_1$ player. $\blacksquare$

Our investigations indicate other Pareto Optimal solutions exist,
but the one in Theorem~\ref{Th3} is known as a social welfare
maximizer or socially optimal solution because it maximizes the
total utility for the population.

If every player chooses to be a governor and untrustworthy, the
net individual wealth is zero. In contrast, if a player chooses to
be in the $x_1$ population, he or she can lose all of their money
if all of the governors are untrustworthy. Would a single
trustworthy governor emerge in this population to maximize the
combined wealth of the population?

The evolutionary behavior of a population playing the
trust game can be studied using replicator
dynamics~\cite{replicator}. Let $y_k$ be the frequency of the $x_k$ players
in an infinitely large population with $\sum_k y_k = 1$. Then the time evolution
of $y_k$ is given by the differential
equation

\begin{equation*}
\dot{y}_{k}= {y_k} \cdot \left( {{f_k} - \bar{f}} \right)
\end{equation*}

\noindent where $f_k$ is the expected fitness of an individual
playing strategy $k$ at time $t$ and $\bar{f}$ is the mean
population fitness. Here, fitness and net wealth are equivalent.
The number of copies of a strategy increases if $f_k>\bar{f}$ and
decreases if $f_k<\bar{f}$. We can calculate $\bar{f}$ as follows

\begin{equation*}
 \bar{f} = \frac{y_1 \cdot y_2 \cdot tv \left( 2 \cdot R1 - 1 \right) + y_1 \cdot y_3 \cdot tv \cdot \left( R2 - 1 \right)}{\left( y_2 + y_3 \right)}
\end{equation*}

\vspace{1ex}

 \noindent The three replicator equations (Note that $y_2 + y_3 = 1- y_1 $) are

\vspace{1ex}


\begin{multline*}
 \dot{y}_{1}= \frac{y^2_1 \cdot tv}{1-y_1} \left( y_2 \left( 1 -
2 \cdot R1 \right) + y_3 \cdot \left( 1 - R2 \right) \right) + \\
 \frac{y_1 \cdot tv}{1-y_1} \left( y_2 \left( R1 - 1 \right) - y_3
 \right)
\end{multline*}

\begin{equation*}
 \dot{y}_{2}= \frac{y_1 \cdot y_2 \cdot tv}{1 - y_1} \cdot \left( y_2 \left( 1 - 2 \cdot R1 \right) + y_3 \left( 1 - R2 \right) + R1\right)
\end{equation*}

\begin{equation*}
 \dot{y}_{3}= \frac{ y_1 \cdot y_3 \cdot tv}{1 - y_1} \cdot \left( y_2 \left( 1 - 2 \cdot R1 \right) + y_3 \left( 1 - R2 \right) + R2\right)
\end{equation*}

\vspace{1ex}

\begin{figure}[!t]
\centering
\includegraphics[width=3.5in,height=2.5in]{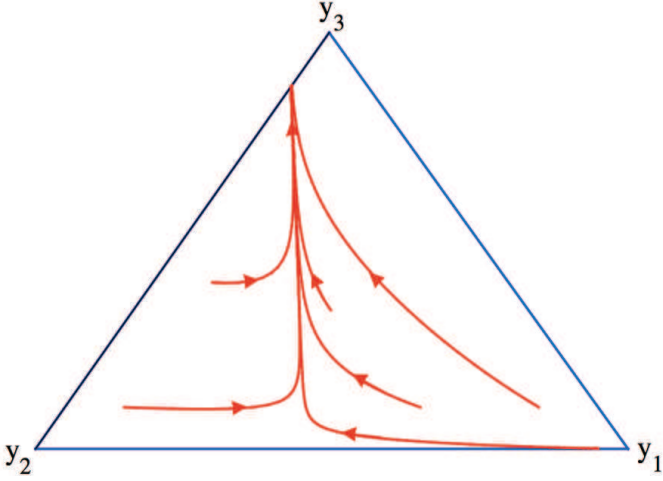}
\caption{A 2-simplex showing the time evolution for a game with $R1=6$, $R2=8$, $tv=10$,
and different initial distributions of $y_1, y_2$ and $y_3$.}
\label{Timeevolution1}
\end{figure}

Figure~\ref{Timeevolution1} shows the population evolution for
various initial player distributions. The replicator equations
predict there is a critical ratio of $x_1$ to $x_2$ players that
acts as an attractor where players rapidly switch strategy to
become untrustworthy. However, untrustworthy players do not
completely take over the population because a certain proportion
always remains trustworthy even when there are a few or no
investors left. Without investors---i.e., $y_1 = 0$---each
replicator equation has the form $\dot{y}_k=0$, which is a fixed
point. Importantly, when there are no longer any investors, the
net worth of all remaining players is zero. Consequently, there is
no incentive to switch strategies and a steady-state condition
exists.

The bottom trajectory in Figure~\ref{Timeevolution1} shows an extreme case that is highly
favorable to a small number of trustworthy players: a population
composed almost entirely of investors with almost no untrustworthy
players. Under these circumstances, the net worth of a trustworthy
player is much higher than that of an investor; therefore
evolution should favor the trustworthy players. Untrustworthy
players have an even higher net worth, but they are initially rare
in the population. The replicator equations initially predict very
rapid growth in trustworthy players with a corresponding plummet
in investors. However, the inevitable increase of untrustworthy
players reverses the trustworthy player growth. Even in this
extreme case, the steady-state condition is reached.

The replicator equations predict interesting behavior in the trust game. Regions to the left
of the attractor have low $C$ to $G$ ratios. Since investments are split among $G$ players
higher returns go to $x_1$ players if most $G$ players are trustworthy. The replicator equations
predict $x_2$ players will mostly switch to $x_1$ players. Conversely to the right of the attractor
there is a high $C$ to $G$ ratio. Under those circumstances there is a strong temptation to be
untrustworthy and the replicator equations predict a sharper rise in $x_3$ players. Near the
attractor the returns to $x_1$ and $x_2$ players are roughly the same so there is little incentive
to switch from $x_2$ to $x_1$ or vica versa. The maximum return is obtained by becoming untrustworthy
and the replicator equations predict virtually all strategy changes are to $x_3$ players. Eventually the
trajectories intersect the $y_2-y_3$ line where there are no investors. At that point all replicator equations
are of the form $\dot{y}_k = 0$ and a fixed point is reached. This makes perfect sense because, without
any investors, there is no incentive to switch strategies.

\begin{figure}[!t]
\centering
\includegraphics[width=3.5in,height=2.5in]{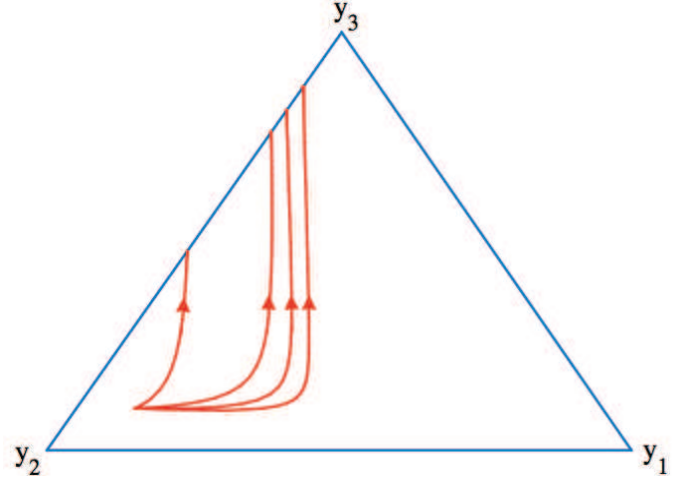}
\caption{A 2-simplex showing the time evolution for a game with $tv = 10$ and different $R1$ and $R2$ values ($R1<R2$). Values increase from left to right with $R1 = 1.5, R2 = 2.9$ for the far left trajectory to $R1 = 6, R2 = 8$ for the far right trajectory. Initial distribution
is $y_1(0)=0.1$, $y_2(0)=0.8$, and
$y_3(0)=0.1$.} \label{Timeevolution4}
\end{figure}

Figure~\ref{Timeevolution4} shows the effect of different
$R1$ and $R2$ values (but with $R1<R2$). A fixed point is still always reached, but the lower the
values, the more trustworthy players in the final population. The reason is
lower $R1$ and  $R2$ values cause the $y_1$ players to go extinct quicker, which limits
the growth of untrustworthy players. Notice the similarity of these trajectories to those
in Figure~\ref{Timeevolution1}, which suggests the presence of an attractor. However, although all attractors are
oriented in the same direction,  their location in the 2-simplex depends on the $R1$ and $R2$ values.

\vspace{4ex}

\begin{theorem}\label{Th4} $y_3 = 0$, $y_1 = \frac{R1-1}{2 R1 - 1}$ and $y_2 = \frac{R1}{2 R1 - 1}$ is a fixed point. \end{theorem}

\vspace{2ex}

\textit{Proof.} The above fixed point is a non-trivial solution
for the replicator equations as a homogeneous
system.$\blacksquare$

\begin{figure}[!t]
\centering
\includegraphics[width=3.5in,height=2.5in]{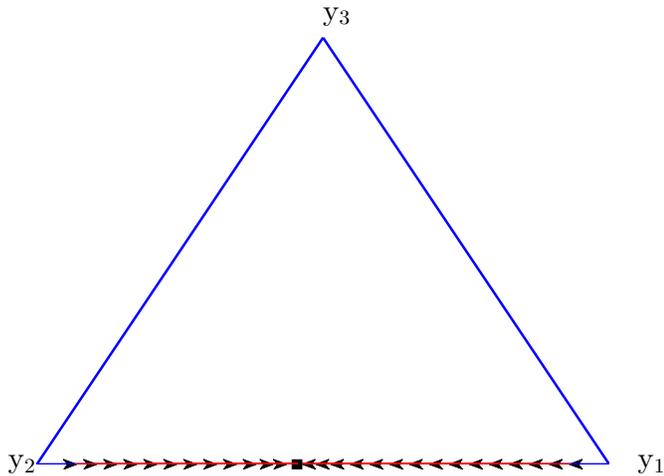}
\caption{A 2-simplex showing the time evolution for a game with
$tv = 10$, $R1=6$, $R2=8$ with different ratios of $y_1$ and $y_2$
while maintaining $y_3=0$ and $y_1+y_2+y_3=1$.}
\label{Timeevolution5}
\end{figure}

Figure~\ref{Timeevolution5} shows the fixed point discussed in
Theorem~\ref{Th4}. Any starting point on the $y_1-y_2$ axis, where
$y_1+y_2+y_3=1$ and $y_3 = 0$ would converge to a fixed point on
this axis. The fixed point in this example, where $R1=6$ is
$y_1=0.454$ and $y_2=0.545$. This fixed point is shown with a
black square in Figure~\ref{Timeevolution5}.

In summary, the replicator equations predict a rapid growth of
untrustworthiness in the population, leading to the eventual
extinction of investors. However, a fraction of the population
always remains trustworthy, even in the absence of investors. This
predicted steady-state outcome is independent of the initial
player distribution, but the ratio of trustworthy to untrustworthy
players in the final population is dependent on the $R1$ and $R2$
values.

The above study and the literature review demonstrate that the
area of modelling trust in evolutionary game theory is in its
infancy. Many opportunities exist to study this fascinating topic.
For example, similar to Yeh and Yang~\cite{yeh2015social}, one can
study the dynamics of the game on a social network, or use genetic
algorithms to evolve strategies for the
game~\cite{david2014genetic}. Many of the assumptions that have
been relaxed for the iterated prisoner dilemma still hold in the
trust games discussed above. For example, one can study the impact
of memory size on the game~\cite{li2014effect} and the impact of
memory on the cognitive resources of the
agents~\cite{ghoneim2013distributing}. Last, but not least,
opportunities exist on the study of the fitness landscape of the
trust game~\cite{he2015easiest,munoz2015exploratory}, which can
reveal insight into the level of hardness or difficulties in
discovering novel strategies, especially in the context of mixed
strategy.

\section{Conclusion}
Trust is a fundamental concept for social systems and individuals
alike. Previous games of trust are limited to 2 players. Previous
work in neuroeconomics introduced basic forms of two-player games
to model trust and described the relationships between trusting
decisions and human neural functions. These games do not create a
social dilemma where the total social wealth depends on the
decisions of both truster and trustee, and they do not generalize
to multiple players. The field of evolutionary computation has
seen little or no work on trust, despite the socioeconomic and
psychological significances of the concept.

In this work, we introduce a new N-player trust game that can
model trust decisions among many players. It creates a social
dilemma in which individuals who attempt to maximize their own
benefits in the short run maximize neither the society's social
wealth nor their own benefits in the long run. The results
revealed that, while the optimal solution for the population
exists when N-1 players choose to be trusters and the remaining
players choose to be trustees, this optimal solution is a needle
in a haystack, causing the evolutionary dynamics to consistently
converge to a population without any trusters and with a
combination of trustworthy and untrustworthy individuals. The
exception to this phenomenon occurs when the initial population is
free of untrustworthy players. The proposed game shows that trust
is an uneasy but worthwhile concept.

\section*{Acknowledgement}
Portions of this work was funded by the Australian Research Council Discovery
Grant number DP140102590.



\begin{thebibliography}{10}
\providecommand{\url}[1]{#1}
\csname url@samestyle\endcsname
\providecommand{\newblock}{\relax}
\providecommand{\bibinfo}[2]{#2}
\providecommand{\BIBentrySTDinterwordspacing}{\spaceskip=0pt\relax}
\providecommand{\BIBentryALTinterwordstretchfactor}{4}
\providecommand{\BIBentryALTinterwordspacing}{\spaceskip=\fontdimen2\font plus
\BIBentryALTinterwordstretchfactor\fontdimen3\font minus
  \fontdimen4\font\relax}
\providecommand{\BIBforeignlanguage}[2]{{%
\expandafter\ifx\csname l@#1\endcsname\relax
\typeout{** WARNING: IEEEtranS.bst: No hyphenation pattern has been}%
\typeout{** loaded for the language `#1'. Using the pattern for}%
\typeout{** the default language instead.}%
\else
\language=\csname l@#1\endcsname
\fi
#2}}
\providecommand{\BIBdecl}{\relax}
\BIBdecl

\bibitem{adolphs2002trust5}
R.~Adolphs, ``Trust in the brain,'' \emph{Nature neuroscience}, vol.~5, no.~3,
  pp. 192--193, 2002.

\bibitem{Ashl1}
D.~Ashlock, E.~Y. Kim, and N.~Leahy, ``Understanding representational
  sensitivity in the iterated prisoner's dilemma with fingerprints,''
  \emph{IEEE Transactions on Systems, Man and Cybernetics - Part C}, vol.~36,
  no.~4, 2006.

\bibitem{baier1986trust7}
A.~Baier, ``Trust and antitrust,'' \emph{Ethics}, pp. 231--260, 1986.

\bibitem{ball2004role8}
D.~Ball, P.~Sim{\~o}es~Coelho, and A.~Mach{\'a}s, ``The role of communication
  and trust in explaining customer loyalty: An extension to the ecsi model,''
  \emph{European Journal of Marketing}, vol.~38, no. 9/10, pp. 1272--1293,
  2004.

\bibitem{brooks2014m10}
A.~W. Brooks, H.~Dai, and M.~E. Schweitzer, ``I?m sorry about the rain!
  superfluous apologies demonstrate empathic concern and increase trust,''
  \emph{Social Psychological and Personality Science}, vol.~5, no.~4, pp.
  467--474, 2014.

\bibitem{darwen1997speciation}
P.~J. Darwen and X.~Yao, ``Speciation as automatic categorical
  modularization,'' \emph{Evolutionary Computation, IEEE Transactions on},
  vol.~1, no.~2, pp. 101--108, 1997.

\bibitem{dasgupta2000trust}
P.~Dasgupta, ``Trust as a commodity,'' \emph{Trust: Making and breaking
  cooperative relations}, vol.~4, pp. 49--72, 2000.

\bibitem{david2014genetic}
O.~E. David, H.~J. van~den Herik, M.~Koppel, and N.~S. Netanyahu, ``Genetic
  algorithms for evolving computer chess programs,'' \emph{Evolutionary
  Computation, IEEE Transactions on}, vol.~18, no.~5, pp. 779--789, 2014.

\bibitem{delgado2005perceptions11}
M.~R. Delgado, R.~H. Frank, and E.~A. Phelps, ``Perceptions of moral character
  modulate the neural systems of reward during the trust game,'' \emph{Nature
  neuroscience}, vol.~8, no.~11, pp. 1611--1618, 2005.

\bibitem{deutsch1962cooperation12}
M.~Deutsch, ``Cooperation and trust: Some theoretical notes.'' in
  \emph{Proceedings of Nebraska Symposium on Motivation}, M.~R. Jones,
  Ed.\hskip 1em plus 0.5em minus 0.4em\relax Univer. Nebraska Press, 1962.

\bibitem{deutsch1977resolution13}
------, \emph{The resolution of conflict: Constructive and destructive
  processes}.\hskip 1em plus 0.5em minus 0.4em\relax Yale University Press,
  1977.

\bibitem{fehr2003nature}
E.~Fehr and U.~Fischbacher, ``The nature of human altruism,'' \emph{Nature},
  vol. 425, no. 6960, pp. 785--791, 2003.

\bibitem{fogel1993evolving}
D.~B. Fogel, ``Evolving behaviors in the iterated prisoner's dilemma,''
  \emph{Evolutionary Computation}, vol.~1, no.~1, pp. 77--97, 1993.

\bibitem{gambetta2000can14}
D.~Gambetta \emph{et~al.}, ``Can we trust trust,'' \emph{Trust: Making and
  breaking cooperative relations}, vol. 2000, pp. 213--237, 2000.

\bibitem{ghoneim2013distributing}
A.~Ghoneim, G.~W. Greenwood, and H.~Abbass, ``Distributing cognitive resources
  in one-against-many strategy games,'' in \emph{Evolutionary Computation
  (CEC), 2013 IEEE Congress on}.\hskip 1em plus 0.5em minus 0.4em\relax IEEE,
  2013, pp. 1387--1394.

\bibitem{giffin1967contribution24}
K.~Giffin, ``The contribution of studies of source credibility to a theory of
  interpersonal trust in the communication process.'' \emph{Psychological
  bulletin}, vol.~68, no.~2, p. 104, 1967.

\bibitem{grodzinsky2014developing19}
F.~S. Grodzinsky, K.~W. Miller, and M.~J. Wolf, ``Developing automated
  deceptions and the impact on trust,'' \emph{Philosophy \& Technology}, pp.
  1--15, 2014.

\bibitem{he2015easiest}
J.~He, T.~Chen, and X.~Yao, ``On the easiest and hardest fitness functions,''
  \emph{Evolutionary Computation, IEEE Transactions on}, vol.~19, no.~2, pp.
  295--305, 2015.

\bibitem{helliwell2011well20}
J.~F. Helliwell and H.~Huang, ``Well-being and trust in the workplace,''
  \emph{Journal of Happiness Studies}, vol.~12, no.~5, pp. 747--767, 2011.

\bibitem{replicator}
J.~Hofbauer and K.~Sigmund, ``Evolutionary game dynamics,'' \emph{Bulletin of
  the American Mathematical Society}, vol.~40, no.~4, pp. 479--519, 2003.

\bibitem{ishibuchi2005evolution}
H.~Ishibuchi and N.~Namikawa, ``Evolution of iterated prisoner's dilemma game
  strategies in structured demes under random pairing in game playing,''
  \emph{Evolutionary Computation, IEEE Transactions on}, vol.~9, no.~6, pp.
  552--561, 2005.

\bibitem{krueger2007neural26}
F.~Krueger, K.~McCabe, J.~Moll, N.~Kriegeskorte, R.~Zahn, M.~Strenziok,
  A.~Heinecke, and J.~Grafman, ``Neural correlates of trust,''
  \emph{Proceedings of the National Academy of Sciences}, vol. 104, no.~50, pp.
  20\,084--20\,089, 2007.

\bibitem{li2014effect}
J.~Li and G.~Kendall, ``The effect of memory size on the evolutionary stability
  of strategies in iterated prisoner's dilemma,'' \emph{Evolutionary
  Computation, IEEE Transactions on}, vol.~18, no.~6, pp. 819--826, 2014.

\bibitem{luhmann1979trust32}
N.~Luhmann, H.~Davis, J.~Raffan, and K.~Rooney, \emph{Trust; and, Power: two
  works by Niklas Luhmann}.\hskip 1em plus 0.5em minus 0.4em\relax Wiley
  Chichester, 1979.

\bibitem{masuda2012coevolution}
N.~Masuda and M.~Nakamura, ``Coevolution of trustful buyers and cooperative
  sellers in the trust game,'' \emph{PloS one}, vol.~7, no.~9, p. e44169, 2012.

\bibitem{mayer1995integrative33}
R.~C. Mayer, J.~H. Davis, and F.~D. Schoorman, ``An integrative model of
  organizational trust,'' \emph{Academy of management review}, vol.~20, no.~3,
  pp. 709--734, 1995.

\bibitem{mittal2009optimal}
S.~Mittal and K.~Deb, ``Optimal strategies of the iterated prisoner's dilemma
  problem for multiple conflicting objectives,'' \emph{Evolutionary
  Computation, IEEE Transactions on}, vol.~13, no.~3, pp. 554--565, 2009.

\bibitem{morrison1997employees36}
E.~W. Morrison and S.~L. Robinson, ``When employees feel betrayed: A model of
  how psychological contract violation develops,'' \emph{Academy of management
  Review}, vol.~22, no.~1, pp. 226--256, 1997.

\bibitem{munoz2015exploratory}
M.~Munoz, M.~Kirley, S.~K. Halgamuge \emph{et~al.}, ``Exploratory landscape
  analysis of continuous space optimization problems using information
  content,'' \emph{Evolutionary Computation, IEEE Transactions on}, vol.~19,
  no.~1, pp. 74--87, 2015.

\bibitem{petraki2014trust0}
E.~Petraki and H.~Abbass, ``On trust and influence: A computational red teaming
  game theoretic perspective,'' in \emph{Computational Intelligence for
  Security and Defense Applications (CISDA), 2014 Seventh IEEE Symposium
  on}.\hskip 1em plus 0.5em minus 0.4em\relax IEEE, 2014, pp. 1--7.

\bibitem{reina2000trust23}
D.~S. Reina and M.~L. Reina, ``Trust and betrayal in the workplace building
  effective relationships in your organization,'' \emph{Advances in Developing
  Human Resources}, vol.~2, no.~1, pp. 121--121, 2000.

\bibitem{spector2004trust41}
M.~D. Spector and G.~E. Jones, ``Trust in the workplace: Factors affecting
  trust formation between team members,'' \emph{The Journal of social
  psychology}, vol. 144, no.~3, pp. 311--321, 2004.

\bibitem{wells2001trust43}
C.~Wells, ``Trust, gender, and race in the workplace,'' \emph{Journal of Social
  Behavior and Personality}, vol.~16, no.~1, pp. 115--126, 2001.

\bibitem{yeh2015social}
C.~Yeh and C.~Yang, ``Social networks and asset price dynamics,''
  \emph{Evolutionary Computation, IEEE Transactions on}, vol.~19, no.~3, pp.
  387--399, 2015.

\end{thebibliography}
\end{document}